\documentclass[11pt]{article}
\usepackage{amsmath,epsf,amssymb,latexsym,enumerate,cite,shadow,array,color}
\usepackage[ps,dvips,matrix,arrow,frame,import,curve,color]{xy}
\DeclareFontFamily{U}{rsf}{}
\DeclareFontShape{U}{rsf}{m}{n}{
  <5> <6> rsfs5 <7> <8> <9> rsfs7 <10-> rsfs10}{}
\DeclareMathAlphabet\Scr{U}{rsf}{m}{n}

\makeatletter
\@addtoreset{equation}{section}
\makeatother

\def\be{\begin{equation}}
\def\ee{\end{equation}}

\def\ba{\begin{array}}
\def\ea{\end{array}}

\newcommand{\bea}{\begin{eqnarray}}
\newcommand{\eea}{\end{eqnarray}}
\def\N{$\cal N$}
\def\E {$E_{7(7)}$}

\begin{document}
\hfill{}

\vskip 1cm

\vspace{24pt}

\begin{center}
{ \LARGE {\bf  On  UV Finiteness of the   Four Loop
\N=8 Supergravity  }}

\vspace{1.5cm}

  {\large  {\bf   Renata Kallosh}}

    \vspace{15pt}

 {Department of Physics,
    Stanford University, Stanford, CA 94305}

\vspace{10pt}

\vspace{24pt}

\end{center}

\begin{abstract}

The 4-loop 4-point amplitude in \N=8 d=4 supergravity is UV finite due to supersymmetry. Even better UV behavior of the 4-loop 4-point amplitude, analogous to that of \N=4 SYM theory,  has been recently established by computation in [1]. 
 All $n$-point 4-loop amplitudes with $n > 5$ are finite on dimensional grounds. However, the situation with the 5-point amplitudes remained unclear. In this paper we will show that  the 5-point 4-loop amplitude must  be finite due to \N=8 supersymmetry, despite the fact that $R^5$ has a supersymmetric generalization for \N=1, \N=2 and  \N=4 SUSY.  This means that
 all 4-loop amplitudes in \N = 8 supergravity are UV finite. We also discuss the current expectations for higher loops.

\end{abstract}
\newpage

\section{Introduction}

Recent {\it tour de force} computations \cite{Bern:2009kd}  of the 4-loop 4-point amplitude in \N=8 d=4 supergravity \cite{Cremmer:1979up} points out towards the possibility of the all-loop UV finiteness of the theory. The purpose of this note is to clarify the  supersymmetry  predictions for the 4-loop \N=8 d=4 supergravity (SG) and comment on higher loop predictions.

The linearized 3-loop counterterm was constructed in \cite{Kallosh:1980fi}, \cite{Howe:1981xy} and for a while it was considered as  a candidate for a 3-loop logarithmic divergence. However, the computations in \cite{Bern:2007hh}  have shown that the corresponding divergence is absent, in agreement with their earlier unitarity cut method expectations. Moreover, not only the term $\log \Lambda \, R^4$, but also ${\partial^2 R^4\over \Lambda^2}$ and ${\partial ^4R^4\over \Lambda^4}$ in d=4 were shown to cancel at the 3-loop level. This ``superfiniteness'' property still does not have a clear explanation, but it indicates that the formula for the critical dimension were the UV divergences start,
\be
D_c= 4+{6\over L}\ ,
\label{YM}\ee
 may be valid in \N=8 SG.  At the 3-loop level in \N=8 SG only the term 
\be
\kappa^4 \int d^4 x \, \sqrt {-g} \, R^4+...\sim \kappa^4 \int d^4 x \, \sqrt {-g} \, (R_{\alpha \beta\gamma \delta} \overline R_{\dot \alpha \dot \beta \dot \gamma \dot \delta})^2+...
 \ee
 could be associated with the logarithmic divergence in graviton amplitudes. Higher powers of curvature, which may have defined an independent higher-point amplitude divergence, are ruled out by dimensional considerations. Therefore, the computation of the 4-point amplitude in  \cite{Bern:2007hh} was sufficient to establish the finiteness of all $n$-point amplitudes at the 3-loop level: The same counterterm responsible for the 4-point divergence (or its absence) is also responsible for the higher point divergence as it is simply a non-linear completion of the 4-point counterterm. Since the 4-point divergence is absent, all higher point amplitudes at 3 loops are also finite.
 
The situation with the 4-loop divergences requires a more detailed discussion. Even prior to the computation of Ref.  \cite{Bern:2009kd}  it was clear that there should not be any logarithmic divergences of the 4-loop 4-point amplitude. However, the authors found much more.  They found that the  superfiniteness in the 4-point amplitude takes place even at the 4-loop level. Thus the mysteries continued to accumulate, which gives an additional encouragement towards further investigation of the possible all-loop UV finiteness of \N=8 d=4 supergravity.

On the other hand, there are no calculations so far of the possible divergences of the 5-point amplitudes  \N=8 d=4 supergravity, without which one cannot be sure of  the full 4-loop finiteness of \N=8 d=4 supergravity. More exactly, the higher point counterterms at 4-loop order $\kappa^6 \int d^4 x \, \sqrt {-g} \,  R^n $ for $n>5$ have positive dimension $2(n-5)$ and do not support logarithmic divergences. The only remaining point to check is
 the 5-point graviton amplitude.\footnote{This issue was raised by A. Tseytlin. We are grateful to
 L. Dixon and Z. Bern who informed us about it. } One may wonder whether the relevant counterterm $\kappa^6 \int d^4 x \, \sqrt {-g} \,  R^5 +...$, which is not a non-linear completion of the 4-point counterterm, is available or forbidden by supersymmetry. We will start here with a review of the known facts on this in the literature.

The recent analysis of supersymmetric counterterms in \cite{Bossard:2009sy} is based on the harmonic superspace construction  in \cite{Drummond:2003ex} (DHHK). It  suggests that   no UV divergences are to be expected at the 4-loop order. This includes the 4-point amplitudes as well as all other higher point amplitudes.  
Since it is not known  whether the list of counterterms in harmonic superspace studied by DHHK in\cite{Drummond:2003ex} includes all possible candidates for \N=8 supersymmetric counterterms\footnote{In what follows we will compare, with the help of P. Howe, our candidates with those studied by DHHK.} , one would like to have an independent information on existence/non-existence of  \N=8 supersymmetrization of the $R^5$ term.
 
 The computation of the 5-point 1-loop type II string amplitude was performed in  \cite{Richards:2008jg} where it was shown that the $R^5$ term is absent. This, by itself, may not be sufficient to prove that the  \N=8 SG in four dimensions will not have a 5-point 4-loop 
UV divergence, however, it makes it rather plausible. Moreover, the tree level computation of the 5-point graviton string amplitude was also performed \cite{DS} and it was shown that various contributions to the  $R^5$ cancel. This tree level answer for the string amplitudes does not suffer from the problem of extra states of string theory versus \N=8 SG \cite{Green:2007zzb}, which may affect the 1-loop computations of \cite{Richards:2008jg}. The fact of cancellation of the tree level $R^5$ term in string theory  \cite{DS}  is therefore, again, suggesting that \N=8 SG at the 4-loop level will not have a 5-point amplitude divergence. Still, the $R^5$ term
 could have been allowed by SUSY and just happen to have the coefficient zero at the tree and 1-loop level in string theory.

In view of all this indications that, most likely,  $R^5$ does not have an \N=8 generalization, a direct \N=8 supersymmetry analysis is  still desirable. If the $R^5$ is disallowed by supersymmetry, this means that the 5-point 4-loop amplitude is free of divergences due to \N=8 supersymmetric Ward identities. This is an unambigous prediction for computations which respect \N=8 supersymmetry. If supersymmetry forbids  the $R^5$ terms, this makes the actual computation not necessary.

In this paper we will show   that in \N=1, \N=2 and \N=4 SG theories one can construct linearized supersymmetric 5-point counterterms starting with $R^5$.  It will be important therefore to study carefully  what exactly is the situation in \N=8.
For this purpose we will evaluate the existence of all possible supersymmetric invariants  following the procedure developed in the past in \cite{Kallosh:1980fi}, \cite{Howe:1981xy} for the 4-point case. We will present the suspects and rule them out case by case.

We will end this note by a short discussion of the possible directions of research of the UV properties of \N=8 SG.

 \section{Analysis of  d=4 4-loop supersymmetric candidate counterterms}

In the 4-loop order no supersymmetric counterterm of the symbolic form $\kappa^6 \int d^4 x \, \sqrt {-g} \, R^4  \partial^2 +...$ is available in d=4, therefore the absence of a logarithmic divergence in the 4-point amplitude is not surprising.

As a warm up consider the supersymmetrization of the 3-point 2-loop $\kappa^2 \int d^4 x R^3$ and 5-point 4-loop graviton coupling $\kappa^6 \int d^4 x R^5$ in \N=1 supergravity.  We can use on shell a chiral conformal superfield $W_{\alpha\beta\gamma} $ of dimension 3/2 and its spinorial derivative $D_{(\delta} W_{\alpha\beta\gamma)}= R_{\alpha\beta\gamma\delta}$. For the 3-point amplitude at 2 loops in d=4 we may try
\be
S_3 \sim \kappa^2 \int d^4x \, d^2 \theta \, W_{\alpha \beta\gamma} \, W^{\gamma\xi\eta} \, W_{\xi \eta }{}^{\alpha} \ .
\ee
It is supersymmetric but has a wrong dimension, so we need  an extra spinorial derivative insertion
\be
S_3 \sim \kappa^2 \int d^4x \, d^2 \theta \, W_{\alpha \beta\gamma} \, W^{\gamma\xi\eta} \, D^{\alpha} W^{\beta}{}_{\xi\eta} \ .
\ee
This term  is not supersymmetric since the insertion of a spinorial derivative makes the superfield $D^{\alpha} W^{\beta}{}_{\xi\eta}$ non-chiral. This is a useful way  to confirm the well known fact that $R^3$ does not have a supersymmetric partner even in \N=1 SG. The 5-point amplitude at 4 loops, however, has an \N=1 supersymmetric  version, namely
\be
S_5^{N=1}  \sim \kappa^6 \int d^4x \, d^2 \theta d^2 \bar \theta\, W_{\alpha \beta\gamma} \, W^{\gamma\xi\eta} \, D^{\alpha} W^{\beta}{}_{\xi\eta} \overline W_{\dot \alpha \dot \beta \dot \gamma} \overline W^{\dot \alpha \dot \beta \dot \gamma} +h.c.
\ee
It corresponds to the following combination of the curvature spinors
\be
\kappa^6 \int d^4x \, R_{\alpha \beta\gamma \delta} \, R^{\gamma\delta \xi \eta} \, R_{\xi\eta}{}^{\alpha \beta}  \overline R_{\dot \alpha \dot \beta \dot \gamma \dot \delta} \overline R^{\dot \alpha \dot \beta \dot \gamma \dot \delta} +h.c.
\label{R5} \ee
In \N=2 supergravity the linearized superfield of dimension 1 is $W_{\alpha \beta}$, which starts with the vector field strength spinor $F_{\alpha \beta}$. The 5-point supersymmetric generalization of the $R^5$ term (\ref{R5}) is
\be
S_5^{N=2} \sim \kappa^6 \int d^4x \, d^4 \theta d^4 \bar \theta\, W_{\alpha \beta} \, D^\beta W^{\gamma\delta} \, D^{\gamma} W^{\delta}{}_{\alpha} \overline W_{\dot \alpha \dot \beta } \, \overline W^{\dot \alpha \dot \beta } +h.c.
\ee
At the level of \N=4 supergravity there is a dimension zero chiral superfield $W$ and the generalization of the
$R^5$ term (\ref{R5}) is
\be
S_5^{N=4} \sim \kappa^6 \int d^4x \, d^8 \theta d^8 \bar \theta\, \epsilon _{ijkl} \, D_\alpha^i D_\beta ^j W \, D^{\alpha k}  W \, D^{\beta l} W \, \overline W \, \overline W +h.c.
\ee

 What is available in \N=8 case?
The full superspace integrals $\kappa^{6} \int d^4 x d^{32} \theta {\cal L}  (W, D, \partial) $ depending on the linearized dimensionless superfield $W_{ijkl}$  and its spinorial and space-time derivatives have positive mass dimension
$>6$ and will not supply the relevant supersymmetric invariant.  
We will  study here the actions over the subspaces of the 32 $\theta$'s.

Thus we would like to look carefully for the counterterms, candidate  for the 4-loop 5-point amplitudes, which are not related to a non-linear completion of the 4-point counterterms, and make sure that all possibilities are taken into account. We will use here the same method  \cite{Kallosh:1980fi}, \cite{Howe:1981xy}  which in the past allowed us not to miss the 3-loop 4-point candidate counterterm.  Now we will apply this method to the 4-loop 5-point case.

We are looking for the  linearized supersymmetric version of $\kappa^6 \int d^4 x \, \sqrt {-g} \,  R^5 +...$.  The linearized superfield of \N=8 supergravity is 
$
W_{ijkl} = {1\over 4!} \epsilon _{ijklmnpr} \overline W^{mnpr}$. 
  We will use here,  for simplicity,  the setting of Ref. \cite{Kallosh:1980fi} where
 the linear superfield $W_{1234} $  depends only on 16 $\theta$'s
 \be
 W\equiv W_{1234} = \overline W^{5678} \equiv  W (x', \theta_B) \ , \qquad  \theta_B = (\theta_1, \theta_2, \theta_3,  \theta_4;  \bar \theta^5,\bar \theta^6, \bar \theta^7, \bar \theta^8)
 \ee
 in a special basis defined in  \cite{Kallosh:1980fi},
$
x' _{\alpha \dot  \alpha}= x_{\alpha \dot  \alpha}+i \sum_1^4\theta_i \sigma_{\alpha \dot  \alpha} \bar \theta^i - i \sum_5^8 \bar \theta^j \sigma_{\alpha \dot  \alpha}  \theta_j 
$.
  The 3-loop 4-point candidate counterterm is
\be
S^{L=3}= \kappa^4 \int d^4 x d^{16}\theta_B W^4 \sim \kappa^4 \int d^4 x \, \sqrt {-g} \, (R_{\alpha \beta\gamma \delta} \overline R_{\dot \alpha \dot \beta \dot \gamma \dot \delta})^2 +...
\ee
Since $W_{1234}^4$ depends only on 16 $\theta_B$, this expression is supersymmetric. Each superfield has a graviton spinor $R_{\alpha\beta\gamma \delta}$ (or $\overline R_{\dot \alpha \dot \beta \dot \gamma \dot \delta}$)  with 4 $\theta$'s (or 4 $\bar \theta$'s). Therefore one of the terms, a 4-graviton part, is invariant under $SU(8)$, so  the supersymmetric partners are also $SU(8)$ invariant. A manifestly $SU(8)$ form of this 3-loop counterterm was constructed in \cite{Howe:1981xy} using the representations theory of $SU(8)$ and the Yang tableaux.

Now we would like to increase the power of $\kappa$ by 2  to describe the 4-loop counterterm.
\be
S^{L=4}_4= \kappa^6 \int d^4 x d^{16}\theta_B W^4 \partial^2 \ .
\ee
Here $W^4 \partial^2$ is a symbolic expression which means that two space time derivatives are inserted between 4 superfields $W^4$. In fact, the action is symmetric in 4 superfields, in the Fourier space we would have
\be
S^{L=4}_4 \sim \delta^4(p_1+p_2+p_3+p_4) W(p_1) W(p_2) W(p_3) W(p_4) (s+t+u) \ .
 \ee
 Since in the 4-point amplitude $s+t+u=0$, there is no 4-loop counterterm supporting the logarithmic divergence.  This explains why the 4-point amplitude at  4-loop order is finite by supersymmetry.

 For the 5-point amplitude we will first identify the supersymmetric invariants and afterwards check their $SU(8)$ invariance. The first indication of the $SU(8)$ invariance will be the presence of the 5-graviton term  (\ref{R5}).

On  dimensional grounds with
$
S^{L=4}_5= \kappa^6 \int d^4 x d^{2m}\theta {\cal L}  (W, D, \partial)
$
we see that $m= 10-\rm dim \, {\cal L} $ where $\rm dim \,  {\cal L}\geq 0$. This means that we have to check the case of 16, 18 and 20 $\theta$-integration with ${\cal L}  (W, D, \partial)$
depending on the linearized dimensionless superfield $W_{ijkl}$  and its spinorial and space-time derivatives. There is no way to have less than 16 $\theta$-integration  since each $W_{ijkl}$ depends at least on 16 $\theta$'s.

The first attempt is\footnote{Such term was considered in DHHK in  \cite{Drummond:2003ex} and ruled out, P. Howe, private communication}
\be
S^{L=4}_5= \kappa^6 \int d^4 x d^{16}\theta_B W^5 \partial^2 \ ,
\label {1}\ee
where $\partial^2$ means that two space-time derivatives are inserted between 5 superfields in an arbitrary way. It looks supersymmetric, since the integrand depends only on 16 $\theta_B$. However, the gravity part of $W$ has 4 $\theta$ or 4 $\bar \theta$, so this expression does not have the 5-graviton part which would be neutral in $SU(8)$. It has, for example, a square of the Bel-Robinson tensor times a scalar field with specific choice of $SU(8)$ indices, in our case $\phi_{1234}$,  
which clearly violates $SU(8)$.

Second attempt\footnote{
This term contains both the $W$  and $\chi$ fields which obey different constraints. 
It has not been studied in an explicit DHHK  analysis in  \cite{Drummond:2003ex}, but can be shown to be not supersymmetric, in agreement with our argument below, P. Howe, private communication. }  is to replace $\partial^2$ by  4 fermionic  derivatives $D_\alpha$, which hit some of the superfields, or to replace one $\partial$ by 2 fermionic derivatives. This has the correct dimension and may have a  5-graviton term, for example:
\be
S^{L=4}_5= \kappa^6 \int d^4 x d^{16}\theta_B W^5 D_{\theta}^4 \ .
\label{ferm}\ee

However, when we hit the superfield $W=W_{1234}$ by a spinorial derivative, say $D_\alpha ^4$,   it becomes a linearized superfield with the first component equal to a spinorial field, a $\mathbf{56} $ of $SU(8)$, namely $\chi_{\alpha 123}$. Consider the properties of this superfield $\chi_{ijk \, \beta}$, which under supersymmetry transforms into the vector field strength $F_{\alpha \beta \,  ij}$ and into the derivative of the scalar $P_{\dot \alpha \beta [ijkl]}$
\be
 D_{\alpha}^k \chi_{ijk \, \beta}= F_{\alpha \beta \,  ij} \ , \qquad  D_{\dot \alpha  l} \chi_{ijk \, \beta}= P_{\dot \alpha \beta [ijkl]} \ .
\ee
 This means that the spinorial  superfield $\chi_{123 \, \beta}$ still depends on $(\theta_1,  \theta_2, \theta_3;  \bar \theta^4, \bar \theta^5,\bar \theta^6, \bar \theta^7, \bar \theta^8)$. However, it does not depend on $\theta_4$ anymore, instead it depends on $\bar \theta^4$. The remaining scalar superfields $W_{1234}$,  which are not hit by the spinorial derivatives (as we have only 4), still depend on the original combination of $\theta$'s, but each of the $\chi$ fields has some of the Grassmann variables switched partially to  the new ones. The integral in eq. (\ref{ferm}) is therefore not supersymmetric.

 The next case is
 \be
S^{L=4}_5= \kappa^6 \int d^4 x d^{16} \theta_B d^2 {\bar \theta^4} W_{1234}^3  \chi_{\alpha 123} \chi ^{\alpha}_{123} + h.c.
\label{ferm1}\ee
This expression looks supersymmetric since the Lagrangian depends on all 18 fermionic directions. However, it is possible to perform the integration over $d^2 {\bar \theta^4}$ since $W_{1234}^3$ does not depend on these fermionic directions. Each of these derivatives will hit only one of the spinorial superfields and produce $\partial_{\alpha \dot \beta} W_{1234}$. The expression becomes equivalent to the one in eq. (\ref{1}) and is, therefore, ruled out.

In case of 20 fermionic integrations, dimension does now permit  any spinorial derivative insertions and the $W_{1234}^5$ terms depends only on 16 $\theta$'s, the integral vanishes, there is no counterterm. We may also try to have a Lagrangian depending on  $W_{1234}^3$ and two other superfields depending on some of $\theta_B$ as well as 4 other $\theta$'s. For example $W_{1235}$ depend on $\theta_1, \theta_2, \theta_3,  \theta_5;  \bar \theta^4,\bar \theta^6, \bar \theta^7, \bar \theta^8$
 \be
S^{L=4}_5= \kappa^6 \int d^4 x d^{16} \theta_B d^2 {\bar \theta^4} d^2 { \theta_5} W_{1234}^3  W_{ 1235} ^2 \ .
\label{20}\ee
This looks supersymmetric, but the 5-graviton term is not there as one can check looking at each superfield  $\theta^4$ and  $\bar \theta^4$ terms \footnote{This type of an invariant was studied in DHHK  \cite{Drummond:2003ex} and ruled out, P. Howe, private communication.}. There are no other sub-superspace integrals depending on any combination of the superfields of the theory with any insertion of superspace derivatives,  which in principle may serve as 5-point 4-loop counterterms.

Thus we conclude that all possibilities to construct the 5-point 4-loop candidate counterterm failed, the amplitude must be finite for the reason of supersymmetry and dimension.

\section{Discussion}

In this paper we have directly established that there is no \N=8 supersymmetric generalization on the $\kappa^6 \int d^4 x \, (R_{....})^5$ counterterm. This is in complete agreement with other indications of the same fact, coming from  \cite{Bossard:2009sy} - \cite{DS}. As the result, there is no need to compute the 5-point 4-loop amplitude in \N=8 SG.

What is in the future for \N=8 SG now that 4-loop amplitudes are established to be finite and even superfinite according to eq. (\ref{YM})? It was pointed out in \cite{Kallosh:2008mq}, \cite{Kallosh:2009db} that the UV properties of \N=8 SG may be studied in the light-cone unconstrained superspace \cite{Brink:1982pd} which admits a set of Feynman rules with one scalar superfield. Only physical degrees of freedom are propagating in this unitary gauge where all local symmetries are fixed. The counterterms for generic L-loop divergences have not been constructed yet in the light-cone formalism. They are known to exist \cite{Howe:1980th}, \cite{Kallosh:1980fi} starting from 8-loop order in terms of the Lorentz covariant on-shell geometric superfields. However,  they may or may not lead to UV divergences. We have seen repeatedly in computations in \cite{Bern:2007hh} and \cite{Bern:2009kd} that the unexplained cancellations may take place.

The analysis performed in \cite{Kallosh:2009db} shows that the relevant linearized counterterms are non-local in the light-cone formalism, which may explain the finiteness of d=4 theory before L=7. When \E \, symmetry
is added to the light-cone analysis, it may lead to the proof of an all loop finiteness of perturbative \N=8 SG. We have shown in \cite{Kallosh:2009db}  that a better understanding  of the structure of the Feynman graphs of the light-cone \N=8 SG may be  useful and may lead to conclusive statements on the puzzling UV properties of the theory. Other proposals suggesting a possibility of UV finiteness of \N=8 SG  \cite{Berkovits:2006vc}, \cite{Green:2007zzb}, \cite{Schnitzer:2007kh}, \cite{ArkaniHamed:2008gz} will likely be clarified and developed in view of the recent impressive computations in \cite{Bern:2009kd}. Hopefully the UV status of perturbative \N=8 SG will be established.

\section*{Acknowledgments}

I thank   Z. Bern,  M. Green, L. Dixon, P. Howe, K. Stelle, S. Stieberger,    A. Tseytlin  and P. Vanhove
 for the discussion of the 5-point amplitudes.
This work is
supported by the NSF grant 0756174.


\end{document}